\documentclass[sigconf,natbib=false]{acmart}

\acmConference[ICCAD '26]{IEEE/ACM International Conference on Computer-Aided Design}{November 8--12, 2026}{San Jose, CA, USA}

\usepackage{amsmath,amsfonts}
\usepackage{algorithmic}
\usepackage{graphicx}
\usepackage{textcomp}
\usepackage{xcolor}
\usepackage{enumitem}
\usepackage{booktabs}
\usepackage{multirow}
\usepackage{gensymb}
\usepackage{flushend}
\usepackage{siunitx}
\usepackage{subcaption}
\usepackage{algorithm}
\usepackage{makecell}
\usepackage{enumitem}

\setcopyright{none}
\renewcommand\footnotetextcopyrightpermission[1]{}

\begin{document}

\title{Structural Verification for Reliable EDA Code Generation without Tool-in-the-Loop Debugging}

\author{Dinithi Jayasuriya}
\affiliation{
  \institution{University of Illinois Chicago}
  \city{Chicago}
  \state{Illinois}
  \country{USA}
}
\email{dkasth2@uic.edu}

\author{Aravind Saravanan}
\affiliation{
  \institution{University of Illinois Chicago}
  \city{Chicago}
  \state{Illinois}
  \country{USA}
}
\email{asara27@uic.edu}

\author{Nilesh Ahuja}
\affiliation{
  \institution{Intel Corporation}
  \country{USA}
}
\email{nilesh.ahuja@intel.com}

\author{Amanda Rios}
\affiliation{
  \institution{Intel Corporation}
  \country{USA}
}
\email{amanda.rios@intel.com}

\author{Amit Trivedi}
\affiliation{
  \institution{University of Illinois Chicago}
  \city{Chicago}
  \state{Illinois}
  \country{USA}
}
\email{amitrt@uic.edu}
\begin{abstract}
Large language models (LLMs) have enabled natural-language-driven automation of electronic design automation (EDA) workflows, but reliable execution of generated scripts remains a fundamental challenge. In LLM-based EDA tasks, failures arise not from syntax errors but from violations of implicit structural dependencies over design objects, including invalid acquisition paths, missing prerequisites, and incompatible API usage. Existing approaches address these failures through tool-in-the-loop debugging, repeatedly executing and repairing programs using runtime feedback. While effective, this paradigm couples correctness to repeated tool invocation, leading to high latency and poor scalability in multi-step settings. We propose to eliminate tool-in-the-loop debugging by enforcing structural correctness prior to execution. Each task is represented as a structural dependency graph that serves as an explicit execution contract, and a verifier-guided synthesis framework enforces this contract through graph-conditioned retrieval, constrained generation, and staged pre-execution verification with diagnosis-driven repair. On single-step tasks, our method improves pass rate from 73.0\% (LLM+RAG) and 76.0\% (tool-in-loop) to 82.5\%, while requiring exactly one tool call per task and reducing total tool calls by more than $2\times$. On multi-step tasks, pass rate improves from 30.0\% to 70.0\%, and further to 84.0\% with trajectory-level reflection. Uncertainty-aware filtering further reduces verifier false positives from 20.0\% to 6.7\% and improves precision from 80.0\% to 93.3\%. These results show that enforcing structural consistency prior to execution decouples correctness from tool interaction, improving both reliability and efficiency in long-horizon EDA code generation.
\end{abstract}

\maketitle

\section{Introduction}

Recent advances in large language models (LLMs) have enabled the translation of natural-language intent into executable programs \cite{jiang2026survey}, opening the possibility of automating complex electronic design automation (EDA) workflows. In EDA, however, tasks require interacting with hierarchical design databases, composing tool-specific APIs, and maintaining consistency across tightly coupled flow stages. While natural-language-driven automation promises to reduce manual scripting effort, reliable execution of AI-generated EDA scripts remains a fundamental challenge.

The core difficulty arises from the structured and stateful nature of EDA environments. Unlike general-purpose code generation, correctness in EDA scripting depends on satisfying implicit dependencies over design objects and operations. Queries and transformations must follow valid object acquisition paths, respect type constraints, and satisfy action-specific preconditions defined by the design hierarchy. As a result, many generated programs fail not due to syntax errors, but due to structurally invalid execution paths, such as missing intermediate objects, invalid parent-child transitions, or incompatible API usage.

Existing LLM-based approaches address this problem primarily through \emph{tool-in-the-loop debugging}, where candidate programs are generated, executed within the EDA tool, and iteratively repaired using runtime feedback. While effective, this paradigm treats the tool as a debugging oracle, leading to repeated execution, high latency, and poor scalability as task complexity increases. More importantly, it does not explicitly model the structural constraints that govern valid execution, relying instead on post-hoc feedback to detect failures. In multi-step query-action settings, where errors propagate across dependent operations, this results in inefficient trial-and-error loops and limited ability to localize root causes.

In this work, we argue that failures in EDA code generation are fundamentally \emph{structural} rather than syntactic, and should be addressed \emph{prior to execution}. Correct execution can be viewed as satisfying an implicit dependency graph over design objects and operations, which encodes valid acquisition paths, multi-branch dependencies, and action-specific constraints. 

Based on this insight, we propose to eliminate tool-in-the-loop debugging by enforcing structural consistency before invoking the external tool. We introduce a framework that extracts a task-level \emph{structural dependency graph} from the natural-language specification and uses it as an explicit execution contract. This graph guides retrieval, constrains code generation, and enables staged pre-execution verification. A multi-layer verifier detects violations such as missing dependencies, invalid object transitions, and API mismatches, while a controller performs diagnosis-driven, localized repair. By shifting error detection and correction to the pre-execution stage, the framework requires only a single tool invocation per task and avoids repeated EDA tool calls.

We evaluate the framework in OpenROAD on both single-step and multi-step query-action tasks over design databases. Compared to LLM+RAG baselines, the proposed method improves pass rate from 73.0\% to 82.5\% on single-step tasks and from 30.0\% to 70.0\% on multi-step tasks. Compared to tool-in-the-loop approaches, it achieves higher accuracy while reducing expensive OpenROAD calls by more than 2$\times$, demonstrating that pre-execution structural verification improves both execution success and efficiency as task complexity increases.

Overall, the paper makes the following contributions:
\begin{itemize}[leftmargin=*, itemsep=2pt, topsep=2pt]
    \item We identify structural dependency violations as a critical source of failure in EDA code generation and formalize execution correctness as consistency with a task-level dependency graph.
    \item We propose a pre-execution verification framework that eliminates reliance on tool-in-the-loop debugging by enforcing structural constraints before execution.
    \item We design a verifier with diagnosis-driven repair that localizes and corrects structural errors without repeated tool interaction.
    \item We demonstrate consistent improvements in correctness and efficiency across OpenROAD tasks, including multi-step workflows, multiple designs, and different PDKs.
\end{itemize}

\section{Related Work}

\vspace{5pt}
\textbf{Agentic Code Generation.}
Recent advances in LLMs have improved code generation, enabling performance on multi-step reasoning, tool use, and long-horizon tasks. Frameworks such as ReAct \cite{yao2022react} and Reflexion \cite{shinn2023reflexion} interleave reasoning and execution for iterative correction through execution traces or self-reflection. Related approaches, including Toolformer \cite{schick2023toolformer}, CodeAct \cite{wang2024executable}, program-aided reasoning \cite{gao2023pal}, and self-debugging \cite{chen2023teaching}, show that structured feedback and tool integration improve correctness. More recent systems extend this paradigm with memory and planning for more complex multi-step behavior. Despite these advances, most approaches still generate candidate programs \cite{park2023generative}, evaluate them via external signals, and iteratively refine them. While effective when execution is cheap, this paradigm relies on post-hoc feedback rather than explicit structural modeling. In structured environments such as EDA, where correctness depends on multi-step dependencies, this leads to inefficient correction.

\vspace{5pt}
\noindent\textbf{LLM-based EDA Scripting and OpenROAD.}
Recent work has explored LLMs for EDA automation, including script generation, design querying, and tool interaction. ChipNeMo \cite{liu2023chipnemo} and RTLCoder \cite{liu2024rtlcoder} show that domain adaptation improves hardware design performance, especially for RTL generation and documentation grounding. EDA Corpus \cite{wu2024eda} provides a public OpenROAD-oriented dataset for reproducible evaluation. Subsequent work focuses on OpenROAD , an open-source physical design system built on OpenDB with stateful stages such as placement, routing, and timing analysis \cite{ajayi2019openroad}. Scripting requires navigating hierarchical design objects and composing tool-specific APIs across design states. OpenROAD-Assistant \cite{sharma2024openroad} and ORAssistant \cite{kaintura2024orassistant} improve API grounding through retrieval-augmented generation but remain prompt-to-code assistants. OpenROAD-Agent adds execution-driven self-correction using runtime feedback, but relies on repeated tool invocation  \cite{wu2025openroad}. In multi-step query-action tasks, this causes repeated execution and overhead as errors propagate across dependent operations.

\vspace{5pt}
\noindent\textbf{Structure-Aware Verification.}
Program correctness in structured environments depends on dependency constraints between operations, not on syntax alone. Prior work in program synthesis uses structured representations \cite{osera2015type}, such as abstract syntax trees, type systems, and syntax-guided synthesis, to constrain generation and reduce invalid outputs. Complementary approaches use generate-then-verify methods \cite{loding2016abstract}, including type checking, constraint solving \cite{wang2018robust}, execution-guided filtering, and \cite{ni2023lever} verifier-based selection . However, these methods generally focus on validating generic properties of generated programs, or depend on post-hoc execution feedback to detect errors, rather than enforcing the domain-specific dependency constraints that govern correct execution before the program is run. They do not capture task-level requirements such as valid object acquisition paths, multi-branch dependencies, and action-specific preconditions in EDA workflows. Feedback-driven methods such as Reflexion \cite{shinn2023reflexion} and self-debugging improve correction, but still depend on global regeneration and repeated execution rather than explicitly isolating structural failures.

\begin{figure}[t]
    \centering
    \includegraphics[width=\columnwidth]{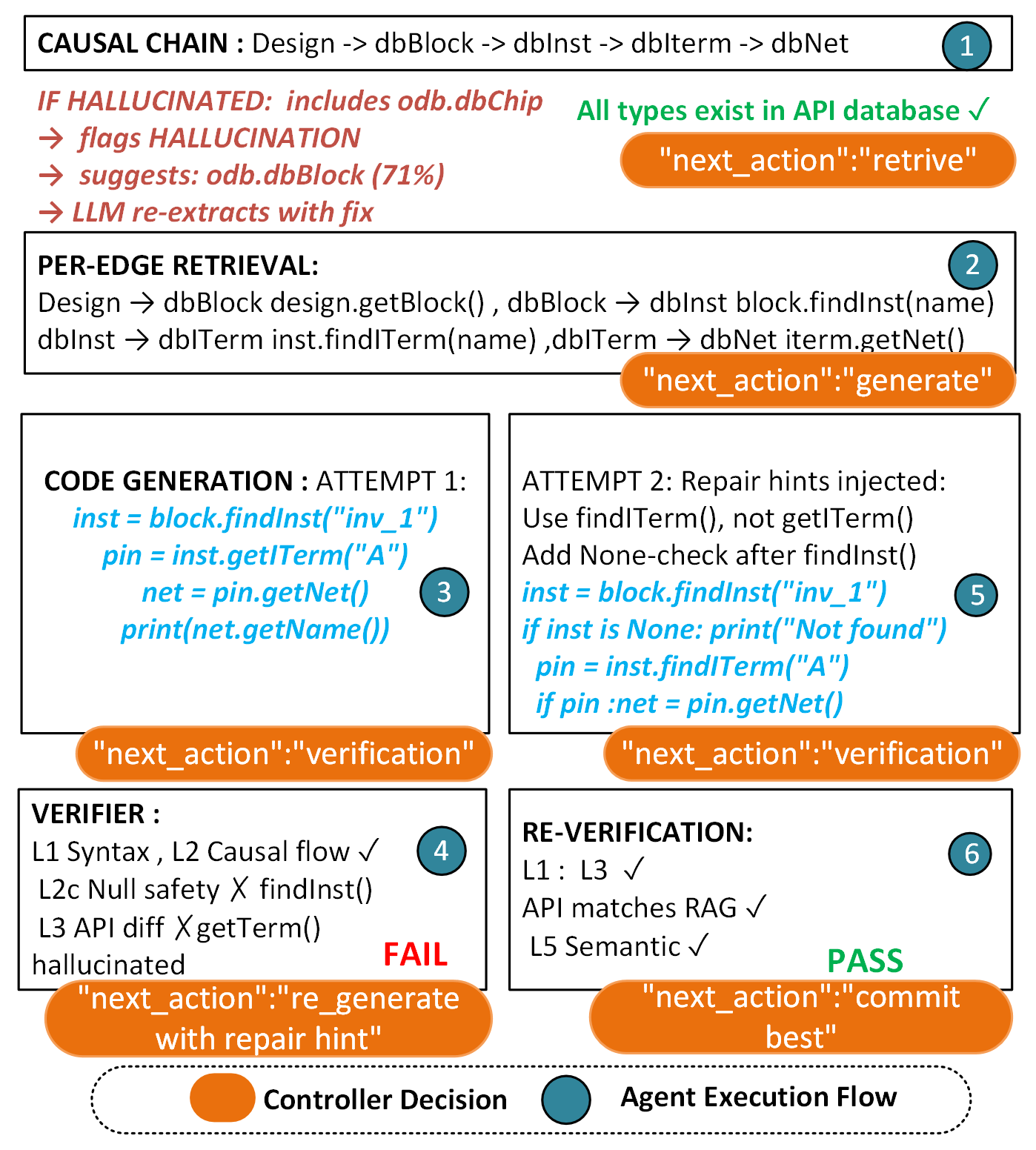}
\caption{Verifier-guided structure-aware synthesis framework with staged verification and localized repair.}\vspace{-10pt}
    \label{causal_graphs}
\end{figure}

In contrast, our approach models EDA tasks as \emph{structural dependency graphs} and enforces structural consistency before execution. Combined with graph-guided generation, staged verification, and localized repair, it avoids tool-in-the-loop debugging and improves efficiency and robustness. 

\section{Verifier-Guided Structure-Aware Synthesis}

\subsection{Methodology Overview}

We address the problem of translating natural-language query-action prompts into executable OpenROAD programs over the OpenDB design database. Given a prompt \( P \), the goal is to generate a program \( \pi \) that executes correctly in a structured, stateful physical design environment. Correctness is governed not only by syntax, but by \emph{structural consistency} with the design hierarchy, including valid object acquisition paths, type compatibility, and action-specific preconditions. We therefore reformulate EDA code generation as enforcing structural consistency \emph{prior to execution}.

In our framework, each task is modeled as a \emph{structural dependency graph} \( G = (V, E) \), where nodes represent typed design objects, intermediate conditions, and actions, and edges encode acquisition and dependency relations \cite{jayasuriya2025causal}. This graph captures required object traversal paths and their composition, and serves as an explicit \emph{execution contract}. A program is valid if it realizes the acquisition paths and satisfies the dependency constraints specified by \( G \). Based on this formulation, we design a verifier-guided synthesis framework that enforces this contract before tool invocation. The framework operates in four stages: (1) graph construction from the prompt, (2) graph-conditioned retrieval and code generation, (3) staged pre-execution verification with diagnosis-driven repair, and (4) extension to multi-step execution with trajectory-level reflection. Unlike execution-driven approaches, all correction occurs prior to tool invocation, requiring a single OpenROAD execution per task. This eliminates reliance on tool-in-the-loop debugging, decouples repair from execution cost, and enables localized correction of structural errors rather than repeated trial-and-error interaction.








\begin{figure}[t]
    \centering
    \includegraphics[width=\columnwidth]{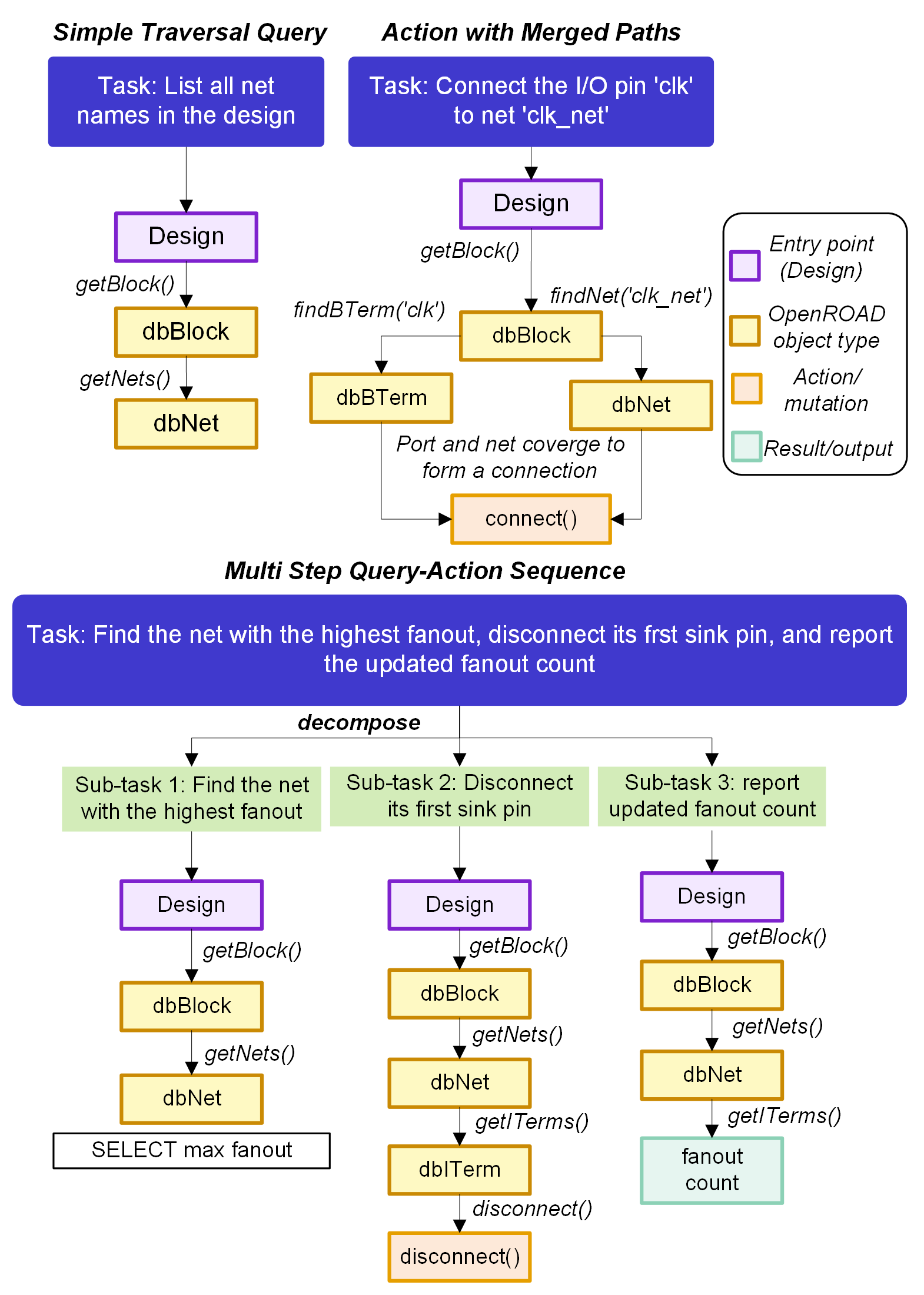}
\caption{
Structural representation of EDA query-action tasks as dependency graphs. \underline{Top:} both simple traversal queries and action queries can be expressed as object-acquisition paths over the design hierarchy, with actions requiring the composition of multiple dependencies. \underline{Bottom:} multi-step tasks decompose into a sequence of such graphs, where each step operates on the updated design state. This formulation exposes the structural constraints underlying execution and motivates enforcing consistency prior to tool invocation.}\vspace{-10pt}
    \label{causal_graphs}
\end{figure}

\subsection{Structural Dependency Graph Construction}

Direct prompt-to-code generation is insufficient because task structure is implicit in natural-language specifications. Even simple prompts require multiple dependent operations, including object acquisition, hierarchical traversal, condition evaluation, and action execution, which are not explicitly enumerated. As a result, correct execution depends on recovering the underlying dependency structure rather than generating code directly.

To make this structure explicit, we construct a \emph{structural dependency graph} \( G = (V, E) \). Nodes represent typed design objects, intermediate targets, conditions, and actions, while edges encode acquisition and dependency relations over the design hierarchy. This representation captures both required objects and their composition into valid execution paths.

The graph supports both linear and multi-branch dependencies. Linear chains correspond to hierarchical traversal (e.g., \texttt{Design $\rightarrow$ Block $\rightarrow$ Net}), while multi-branch structures capture independent object acquisition followed by convergence at an action node (e.g., \texttt{connect(Port, Net)}), avoiding spurious sequential dependencies.

We construct \( G \) via a hypothesis-and-validation process. An initial graph is proposed through structured LLM extraction, producing typed acquisition paths and an action node, which are deterministically parsed into graph form. The extracted graph is then validated against a structured OpenROAD API schema defining valid object types and permissible relations.

Each node and edge must satisfy type and relation constraints defined by the schema. Nodes are classified as valid, missing-but-real, or hallucinated, and edges are checked for valid parent-child transitions. Violations are converted into structured feedback and used to iteratively refine the graph until all constraints are satisfied. This step acts as a structural grounding gate, ensuring that only API-consistent graphs are passed to synthesis. By enforcing validity at the level of object dependencies rather than execution, the resulting graph serves as a reliable execution contract for downstream stages.

\subsection{Structure-Grounded Synthesis}

Given a validated structural dependency graph \( G \), we perform structure-grounded synthesis, where both retrieval and generation are conditioned on the explicit dependency structure rather than the raw prompt. This shifts synthesis from unconstrained sequence generation to structured composition aligned with execution.

\vspace{2pt}
\noindent\textbf{Graph-Conditioned Retrieval.}
Retrieval is decomposed into localized queries aligned with edges and substructures in \( G \). Each query corresponds to an object transition or action dependency, retrieving API usage patterns and code fragments relevant to that relation. This yields evidence tied to individual dependencies, improving grounding while covering implicit prerequisite operations. Aligning retrieval with the graph reduces irrelevant context and helps recover missing intermediate steps.

\vspace{2pt}
\noindent\textbf{Constrained Code Generation.}
Code generation is conditioned on the prompt, the dependency graph \( G \), and the retrieved evidence. The graph enforces structural constraints, restricting generation to valid object-acquisition paths and dependency compositions, while retrieved evidence grounds API usage at each step. This reduces the search space by eliminating structurally invalid programs, such as those with missing prerequisites, incorrect traversal order, or incompatible object types. In contrast to prompt-only generation, where structure must be inferred, the graph provides an explicit plan that guides composition.

The resulting program aligns with the structural execution contract defined by \( G \), but may still contain residual inconsistencies in implementation or API usage. It is subsequently passed to the verifier-guided control stage for validation and targeted repair.

\subsection{Verifier-Guided Control}

This stage closes the synthesis loop by coupling generation with structured feedback and targeted repair, ensuring that the program satisfies the structural execution contract defined by \( G \).

\vspace{4pt}
\noindent\textbf{Four-Layer Verification.}
Each candidate is evaluated through a four-layer pipeline targeting progressively deeper error classes. We denote the failure layer as $\ell \in \{0,1,2,3,4\}$, where $\ell = 0$ indicates all layers are passed. Layer~1 (Syntax) ensures well-formed Python via AST parsing. Layers~2 and~3 perform \textbf{causal verification} against \( G \). Layer~2 (Causal Flow) verifies correct acquisition order, valid attribute access, and proper handling of nullable returns. Layer~3 (API Alignment) checks that invoked methods match retrieved APIs, detecting hallucinated calls and incorrect object usage. Layer~4 (Semantic) evaluates task-level consistency using an LLM judge, including completeness, control flow, and output validity, and is applied only after L1--L3 pass.

\vspace{4pt}
\noindent\textbf{Diagnosis-Driven Repair.}
The verifier produces a structured diagnostic report specifying failure stage, issues, and localized inconsistencies (e.g., missing acquisition steps, invalid transitions, receiver mismatches, incorrect APIs). Each failure is mapped to a region of \( G \), enabling localized correction rather than global regeneration. A controller selects actions based on the task, graph, retrieved APIs, prior actions, feedback, and remaining budget: \textbf{Regeneration} (regenerate code with targeted repair hints for L1, L2, or persistent L3 failures), \textbf{Edge re-retrieval} (update retrieval for a specific graph edge and regenerate for L3 failures), \textbf{Graph re-extraction} (revise \( G \) and restart synthesis for repeated L2 failures), and \textbf{Accept} (commit when minor issues remain but the program is semantically valid). A deterministic safeguard prevents repeated ineffective actions by detecting loops and escalating to alternative strategies.

\vspace{2pt}
\noindent\textbf{Execution Efficiency.}
All correction occurs prior to tool invocation, requiring exactly one OpenROAD call per task. This decouples repair from execution cost and avoids repeated tool interaction. Increasing the repair budget therefore improves solution quality without increasing tool usage, unlike execution-driven approaches where each iteration incurs an additional call.

\subsection{Post-Verification Uncertainty Estimation}
\label{sec:uncertainty}

The verifier enforces structural correctness but produces a binary decision. In practice, some verifier-\textsc{Pass} programs still fail in OpenROAD due to unsupported API usage or unstable repair trajectories. This highlights a limitation of purely structural verification: passing the verifier is necessary but not always sufficient for reliable execution \cite{kumar2025calibrated}. We therefore introduce a post-verification uncertainty estimate that scores the reliability of verifier-\textsc{Pass} programs prior to execution \cite{tayebati2026tracer}. We decompose uncertainty into three sources: \emph{code-level uncertainty}, capturing suspicious API usage in the final program; \emph{trajectory-level uncertainty}, capturing instability in the repair process; and \emph{verification-coverage uncertainty}, capturing how much of the program is actually validated. Let $\mathbf{x}$ denote the final program and $\mathcal{T}$ its repair trajectory:
\(
u(\mathbf{x}, \mathcal{T}) = \alpha_1 u_{\mathrm{code}}(\mathbf{x}) + \alpha_2 u_{\mathrm{traj}}(\mathbf{x}) + \alpha_3 u_{\mathrm{cov}}(\mathcal{T}),
\)
where $\alpha_1 + \alpha_2 + \alpha_3 = 1$.
\subsubsection{Code-Level Uncertainty}

Code-level uncertainty targets residual failures where structurally valid programs reference unsupported API constructs. Let $\mathcal{A}_m$ and $\mathcal{A}_t$ denote valid method and type sets. We define three hallucination signals:

\begin{equation}
\begin{aligned}
h_m(\mathbf{x}) \; &\textbf{(method hallucination)} \;=\;
\frac{\left|\{m \in M(\mathbf{x}) : m \notin \mathcal{A}_m\}\right|}{|M(\mathbf{x})|}, \\
h_I(\mathbf{x}) \; &\textbf{(invalid imports)} \;=\;
\left|\{i \in I(\mathbf{x}) : i \notin \mathcal{A}_m \cup \mathcal{A}_t\}\right|, \\
h_E(\mathbf{x}) \; &\textbf{(enum hallucination)} \;=\;
\left|\{e \in E(\mathbf{x}) : e \notin \mathcal{E}_{\mathrm{known}}\}\right|.
\end{aligned}
\end{equation}

These are combined into a confidence score
\begin{equation}
c_{\mathrm{code}}(\mathbf{x})=
\mathrm{clip}\!\left[
1-\lambda_I h_I(\mathbf{x})
-\lambda_E h_E(\mathbf{x})
-\lambda_r h_m(\mathbf{x}),
\,0,\,1
\right],
\end{equation}
and uncertainty is defined as \(
u_{\mathrm{code}}(\mathbf{x}) = 1 - c_{\mathrm{code}}(\mathbf{x}) \).

\subsubsection{Trajectory-Level Uncertainty}

Trajectory-level uncertainty captures instability in the repair process. Let $\ell_0,\dots,\ell_T$ denote verifier outcomes ($\ell_t \in \{0,\dots,4\}$) and $\mathbf{x}_0,\dots,\mathbf{x}_T$ the candidates.

\begin{equation}
\begin{aligned}
\tau_{\mathrm{conv}} \; &\textbf{(convergence)} \;=\;
\max\!\left(0,\,
1 - \frac{\ell_0 - \ell_T}{\max(\ell_0, 1)}
\right), \\
\tau_{\mathrm{stag}} \; &\textbf{(stagnation)} \;=\;
\frac{1}{T}\sum_{t=0}^{T-1} J(\mathbf{x}_t,\mathbf{x}_{t+1}), \\
\tau_{\mathrm{eff}} \; &\textbf{(repair ineffectiveness)} \;=\;
\frac{\left|\{t \in R : \ell_t \geq \ell_{t-1}\}\right|}{|R|}.
\end{aligned}
\end{equation}

We combine these as
\begin{equation}
u_{\mathrm{traj}}(\mathcal{T})=
w_c \tau_{\mathrm{conv}}
+w_s \tau_{\mathrm{stag}}
+w_e \tau_{\mathrm{eff}}.
\end{equation}

\subsubsection{Verification-Coverage Uncertainty}

Verification-coverage uncertainty captures cases where a verifier \textsc{Pass} reflects incomplete checking rather than confirmed correctness. Let $\mathcal{K}(\tau)$ denote the set of known methods for receiver type $\tau$, and let $C(\mathbf{x})$ be the set of method calls in program $\mathbf{x}$, with receiver types inferred via lightweight propagation. We define two coverage signals:
\begin{equation}
\begin{aligned}
c_{\mathrm{cov}}(\mathbf{x}) \; &\textbf{(verified-call coverage)} \;=\;
\frac{|C_{\mathrm{cov}}(\mathbf{x})|}{|C(\mathbf{x})|}, \\
u_{\mathrm{unc}}(\mathbf{x}) \; &\textbf{(unchecked-call ratio)} \;=\;
\frac{|C_{\mathrm{unc}}(\mathbf{x})|}{|C(\mathbf{x})|},
\end{aligned}
\end{equation}
where
\begin{equation}
\begin{aligned}
C_{\mathrm{cov}}(\mathbf{x}) &= \{(\tau,m)\in C(\mathbf{x}) : \tau \text{ is well-covered and } m \in \mathcal{K}(\tau)\}, \\
C_{\mathrm{unc}}(\mathbf{x}) &= C(\mathbf{x}) \setminus C_{\mathrm{cov}}(\mathbf{x}).
\end{aligned}
\end{equation}

Since these quantities are complementary, we define
\begin{equation}
u_{\mathrm{cov}}(\mathbf{x}) = 1 - c_{\mathrm{cov}}(\mathbf{x}) = \frac{|C_{\mathrm{unc}}(\mathbf{x})|}{|C(\mathbf{x})|}.
\end{equation}
$u_{\mathrm{code}}$, $u_{\mathrm{traj}}$, and $u_{\mathrm{cov}}$ capture complementary risks where unsupported API behavior, unverifiable API usage, and weak convergence, providing a more informative estimate than the verifier's binary verdict and enabling uncertainty-aware filtering prior to execution.

\subsection{Extension to Multi-Step Tasks}

We extend the framework to multi-step query-action prompts by decomposing each input into an ordered sequence of subtasks. Each subtask follows the same pipeline,
\(
\text{subtask} \rightarrow \text{graph} \rightarrow \text{validation} \rightarrow \text{retrieval} \rightarrow \text{generation} \rightarrow \text{verification},
\)
and is executed only after successful verification. Accepted programs update the design state, which is propagated to subsequent steps.

A key challenge is cascading failure, where errors in early steps corrupt the design state and cause downstream failures, even when later steps are locally correct. Thus, correctness must be enforced both locally and across the full execution sequence. If a subtask fails after exhausting the repair budget, execution is terminated to prevent propagation of corrupted state.

To address this, we introduce an episode-level reflection mechanism for cross-step failure diagnosis. After an initial pass, the framework analyzes the full sequence of subtask outcomes to identify root causes of failure. Reflection captures cross-step dependencies and error propagation not observable within individual subtasks, including inconsistencies between earlier object acquisition and later usage. The resulting diagnostic signal guides a second-pass synthesis with targeted corrections, enabling recovery from failures that cannot be resolved through local verification alone. This pass corrects upstream structural errors rather than repeatedly repairing downstream symptoms \cite{madaan2023self}.

\begin{table}[t]
\centering
\caption{Single-step query-action performance and efficiency. Higher pass rate, lower latency and fewer tool calls are better.\vspace{-5pt}}
\label{tab:single_main}
\small
\setlength{\tabcolsep}{1pt}
\begin{tabular}{lcccc}
\toprule
Method & Pass Rate (\%) & Latency (s) & Calls / Task & Total Calls \\
\midrule
Static verifier only        & 50.8 & 19.0 & 1.00 & 120 \\
Causal verifier only        & 79.2 & 27.5 & 1.00 & 120 \\
\midrule
\textbf{Full pipeline}      & \textbf{82.5} & 34.8 & \textbf{1.00} & \textbf{120} \\
\midrule
Tool-in-loop (GPT-4.1-mini) & 76.0 & 53.0 & 1.77 & 248 \\
OpenROAD Agent (Qwen 7B)    & 16.4 & 70.0 & 3.54 & 496 \\
\bottomrule\vspace{-20pt}
\end{tabular}
\end{table}

\begin{table*}[t]
\centering
\caption{OpenROAD script generation performance across systems. We report pass rate, tool usage, and latency.}
\label{tab:openroad_pass_rate_all_systems}
\small
\setlength{\tabcolsep}{4pt}
\begin{tabular}{l l l c c c c c}
\toprule
Model & Method & Generator & Verifier & RAG & Pass (\%) & Calls / Task & Latency (s) \\
\midrule

Qwen2.5-7B 
& OpenROAD-Agent 
& Qwen2.5-ft 
& --- 
& \checkmark 
& 16.4 
& 3.54 
& 69.5 \\

Qwen2.5-7B 
& OpenROAD-Agent + verifier 
& Qwen2.5-ft 
& GPT-4.1-mini 
& \checkmark 
& 31.8 
& 1.00 
& 33.4 \\

\midrule

GPT-4.1-mini 
& LLM only 
& GPT-4.1-mini 
& --- 
& $\times$ 
& 48.0 
& --- 
& --- \\

GPT-4.1-mini 
& LLM + RAG 
& GPT-4.1-mini 
& --- 
& \checkmark 
& 73.0 
& --- 
& --- \\

GPT-4.1-mini 
& LLM + tool-in-loop 
& GPT-4.1-mini 
& --- 
& \checkmark 
& 76.0 
& 1.77 
& 37.6 \\

GPT-4.1-mini 
& \textbf{Ours (full)} 
& GPT-4.1-mini 
& GPT-4.1-mini 
& \checkmark 
& \textbf{82.5} 
& \textbf{1.00} 
& 34.8 \\

\midrule

GPT-4o 
& LLM only 
& GPT-4o 
& --- 
& $\times$ 
& 52.0 
& --- 
& --- \\

GPT-4o 
& \textbf{Ours (full)} 
& GPT-4o 
& GPT-4o 
& \checkmark 
& \textbf{83.0} 
& 1.00 
& 36.2 \\

\bottomrule
\end{tabular}
\end{table*}

\section{Results}

\subsection{Benchmark and Evaluation Protocol}

We evaluate the proposed framework on generating executable OpenROAD programs from natural-language query-action prompts over the OpenDB design database. The evaluation focuses on (i) execution correctness and (ii) efficiency in terms of OpenROAD tool usage. We consider two task settings. \textit{First}, single-step query-action prompts, where each prompt specifies a compact OpenROAD operation. These provide a controlled setting for evaluating generation quality and verifier effectiveness, although correct execution typically requires multiple dependent API calls over the design hierarchy. \textit{Second}, multi-step query-action prompts, where each prompt consists of a sequence of dependent subtasks executed within a single OpenROAD session. This setting is more challenging, as correctness depends on both step-level validity and consistency of the evolving design state. Since no standard benchmark exists, we construct a dataset of 100 complex prompts via adversarial self-play \cite{lin2025learning}, covering diverse object hierarchies, API interactions, and multi-stage operations \cite{cheng2024self}.

A task is considered successful if the generated code executes in the OpenROAD Python shell without raising an exception or traceback. Execution is performed in a live OpenROAD process via a pseudo-terminal. A task is marked as failed if a traceback occurs, if the code contains a syntax error, if execution exceeds 120 seconds, or if the process crashes. While execution success does not guarantee semantic optimality, it is a necessary condition for usable EDA scripts. In OpenROAD, incorrect API usage typically manifests as runtime exceptions, making execution success a strong proxy for correctness. Structural pre-execution verification and post-verification uncertainty filtering mitigate residual risks.

For multi-step tasks, subtasks are executed sequentially within a single session, preserving verified state across steps. A task is successful only if all subtasks execute successfully; any failure terminates execution to prevent error propagation. In addition to correctness, we measure efficiency via OpenROAD tool calls and end-to-end latency, enabling direct comparison with execution-driven approaches where each repair iteration incurs an additional tool call. We also evaluate generalization across designs, PDKs, and OpenROAD stages, e.g. floorplanning and placement.

\subsection{Key Results on Single-Step Tasks}

We evaluate the proposed framework on single-step query-action prompts, a controlled setting for assessing execution correctness. Although each prompt specifies a single operation, correct execution typically requires multiple dependent API calls over the OpenDB hierarchy. Table~\ref{tab:single_main} summarizes the results. The full pipeline achieves a pass rate of \textbf{82.5\%}, outperforming LLM+RAG (73.0\%) and tool-in-the-loop correction (76.0\%), showing that enforcing structural correctness prior to execution improves reliability \cite{lewis2020retrieval}.

To isolate component contributions, we evaluate ablated variants. The static verifier achieves 50.8\% pass rate, while the causal verifier improves performance to 79.2\%, indicating that enforcing object-acquisition structure provides the dominant gain. The full pipeline further increases performance to 82.5\%, showing that semantic validation and repair provide smaller, incremental improvements. Compared to prior systems, the gains are substantial: OpenROAD Agent achieves 16.4\% pass rate, and only 31.8\% even with a verifier, highlighting the importance of integrating structural verification during generation rather than applying it post hoc. Table~\ref{tab:openroad_pass_rate_all_systems} shows consistent trends across models. For GPT-4.1-mini, retrieval improves pass rate from 48.0\% to 73.0\%, and tool-in-the-loop correction to 76.0\%, while our method reaches 82.5\%. Similar gains hold for GPT-4o (52\% to 83\%), demonstrating robustness across models. These results indicate that retrieval alone is insufficient, as it does not enforce structural consistency required for correct execution.

\begin{figure}[t]
    \centering
    \includegraphics[width=\columnwidth]{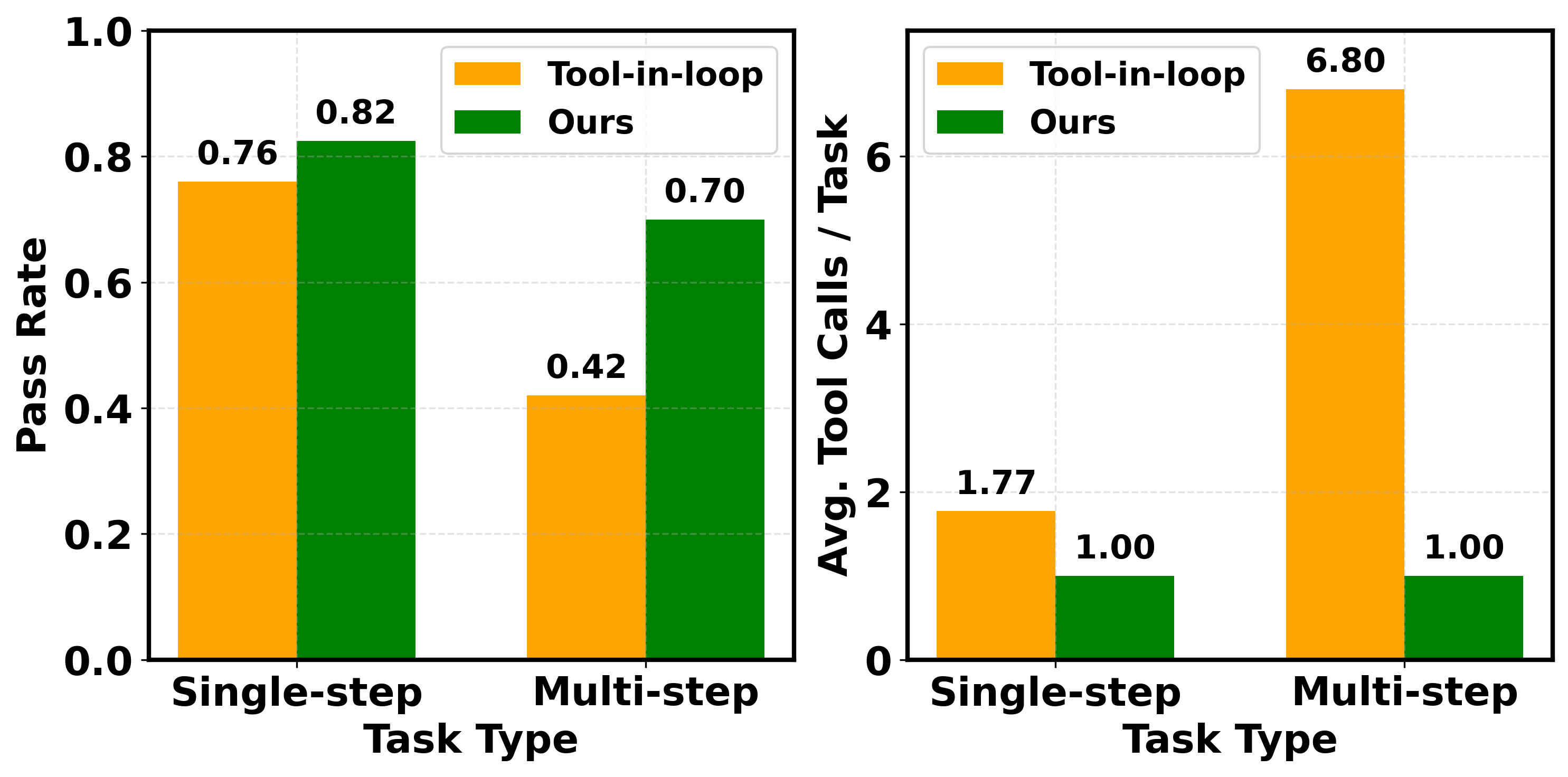}
    \caption{Pass rate and tool usage comparison across single-step and multi-step tasks. The proposed framework maintains higher accuracy while requiring a constant single tool call per task, whereas tool-in-the-loop methods incur significantly higher tool usage and degrade in multi-step settings.}
    \label{fig:efficiency_summary}
\end{figure}

\subsection{Efficiency and Tool Call Reduction}

Table~\ref{tab:single_main} shows that the proposed framework requires exactly \textbf{one OpenROAD call per task}, compared to 1.77 for tool-in-the-loop correction and 3.54 for OpenROAD Agent. Across the evaluation set, this corresponds to \textbf{120 total calls} versus 248 and 496, yielding more than a $2\times$ reduction in tool usage. Latency follows a similar trend. By performing all correction prior to invocation, the framework executes each task only once. Although the full pipeline introduces modest overhead from verification and repair (34.8s vs 27.5s for causal-only), it remains significantly faster than tool-in-the-loop correction (53.0s) and OpenROAD Agent (70s).

\begin{figure}[t]
    \centering
    \includegraphics[width=\linewidth]{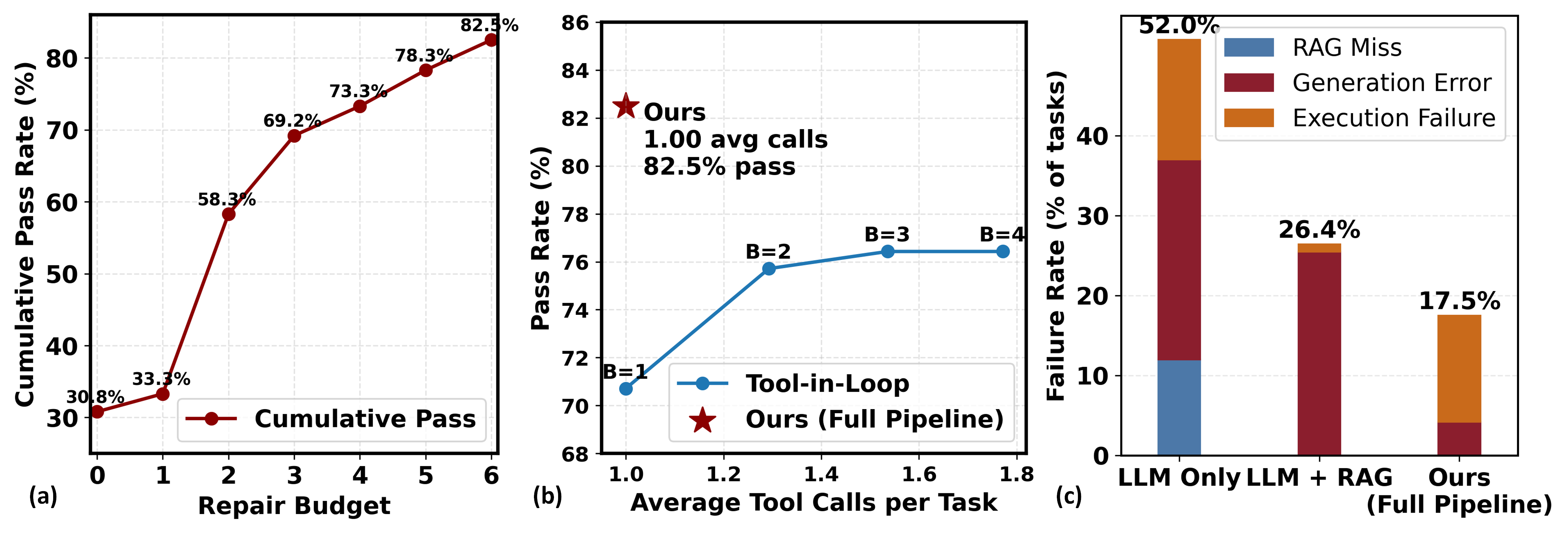}
    \caption{Effect of repair budget and tool interaction on execution performance. 
(a) Cumulative pass rate improves with repair budget, reflecting gains from iterative refinement. 
(b) Pass rate versus average tool calls per task: the proposed framework achieves higher accuracy (82.5\%) with a single tool call, while tool-in-the-loop methods require multiple executions for smaller gains. 
(c) Failure mode breakdown highlighting reductions in generation and execution errors.}
    \label{fig:repair_budget}
\end{figure}

Fig.~\ref{fig:efficiency_summary} compares pass rate and tool usage across single-step and multi-step settings. The proposed method maintains higher accuracy with a constant single tool call per task, while tool-in-the-loop approaches degrade in multi-step settings and incur substantially higher tool usage (up to 6.8 calls per task). This shows that execution cost scales with task complexity for tool-in-the-loop methods but remains fixed under pre-execution verification. Fig.~\ref{fig:repair_budget} further shows that increasing the repair budget improves pass rate without increasing tool usage. In contrast, tool-in-the-loop methods incur additional tool calls with each repair iteration. 

\begin{table}[t]
\centering
\caption{Multi-step query--action performance.}
\label{tab:multi_step_main}
\small
\setlength{\tabcolsep}{4pt}
\begin{tabular}{lcccc}
\toprule
Method & Verifier & RAG & Pass (\%) & Latency (s) \\
\midrule
LLM + RAG        & ---         & \checkmark & 30.0 & 12.2 \\
Ours             & \checkmark  & \checkmark & 70.0 & 53.9 \\
Ours + Reflection& \checkmark  & \checkmark & \textbf{84.0} & \textasciitilde94.9 \\
\bottomrule\vspace{-20pt}
\end{tabular}
\end{table}

\begin{figure*}[t]
    \centering
    \includegraphics[width=1\textwidth]{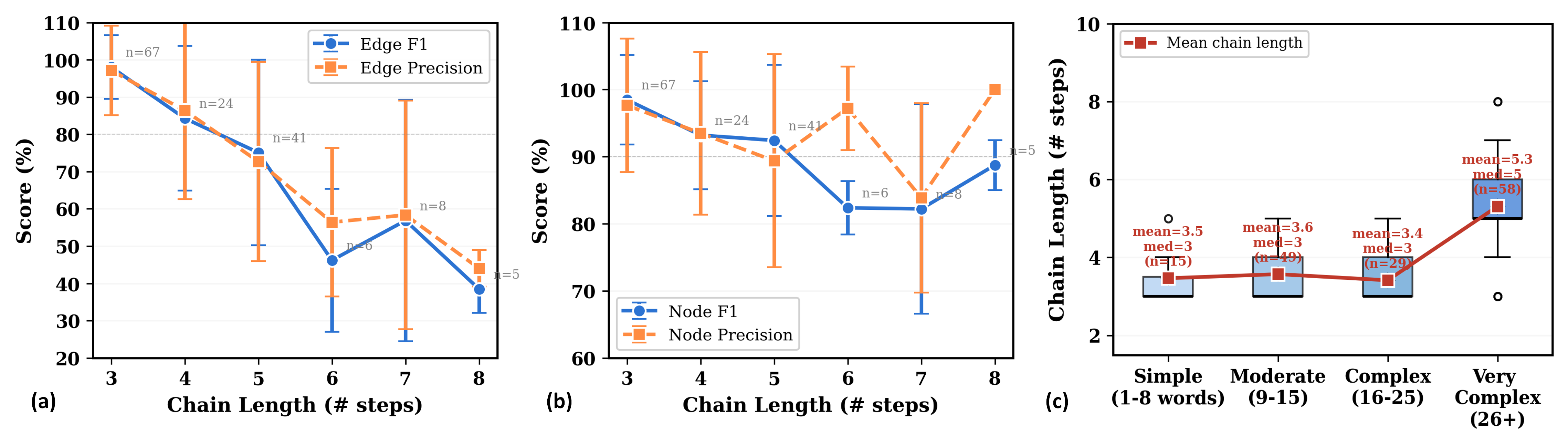}
    \caption{Causal-graph extraction accuracy. (a) Edge-level F1 and precision as a function of chain length. (b) Node-level F1 and precision as a function of chain length. (c) Ground-truth chain length across prompt complexity buckets.}
    \label{causal_graph_acc}
\end{figure*}

\begin{figure*}[t]
    \centering
    \includegraphics[width=1\textwidth]{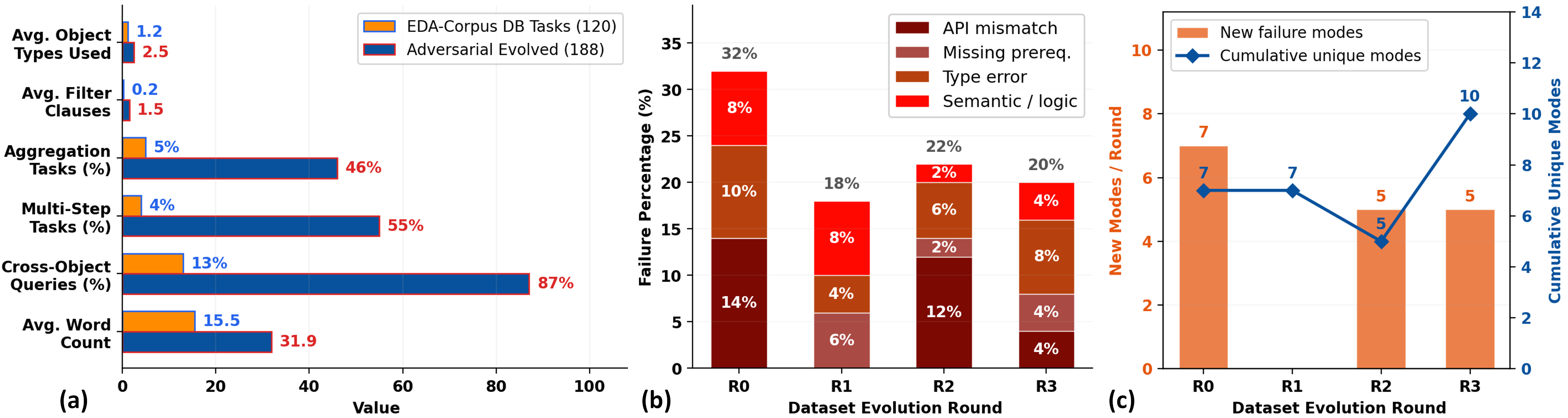}
    \caption{Round-based adversarial evolution for multi-step benchmark construction. (a) Compared to base EDA-Corpus tasks, evolved prompts exhibit greater structural complexity, including multi-step composition, aggregation, and cross-object interactions. (b) Failure types observed across evolution rounds, including API mismatches, missing prerequisites, type inconsistencies, and semantic errors. (c) Newly discovered and cumulative failure modes across rounds, showing that later rounds introduce more diverse and challenging tasks rather than repeating a fixed error distribution.}
    \label{fig:dataset_evolution}
\end{figure*}

\subsection{Multi-Step Execution Performance}

We evaluate the proposed framework on multi-step query-action prompts, where each task consists of dependent subtasks executed within a single OpenROAD session. This setting is substantially more challenging than single-step tasks, as correctness depends on both step-level validity and consistency of the evolving design state. Since no standard benchmark exists for this regime, we construct a multi-step dataset through round-based adversarial evolution, where a creator model increases prompt complexity and a solver model improves solvability across rounds. Figure~\ref{fig:dataset_evolution} summarizes this process. Compared to base EDA-Corpus tasks, the evolved prompts exhibit greater structural complexity and expose a broader range of failure types, including API mismatches, missing prerequisites, type inconsistencies, and semantic failures. Fig.~\ref{fig:dataset_evolution}(c) further shows that new failure modes continue to emerge across rounds, indicating increasing task diversity rather than overfitting to a fixed error distribution.

Table~\ref{tab:multi_step_main} summarizes multi-step execution results on this benchmark. LLM+RAG achieves a pass rate of 30.0\%, reflecting failures due to missing dependencies and inconsistent state propagation. Enforcing structural correctness at each step improves performance to 70.0\%, indicating that most local errors can be resolved through per-step verification even as task complexity increases.Incorporating trajectory-level reflection further improves pass rate from 70.0\% to 84.0\%. This shows that many of the remaining failures arise not from local step errors, but from inconsistencies that propagate across steps. By reasoning over the full trajectory, reflection helps identify and correct these upstream errors.

\begin{table}[t]
\centering
\small
\caption{Causal-chain accuracy by prompt complexity \vspace{-10pt}}
\label{tab:complexity_accuracy}
\begin{tabular}{lcccc}
\hline
\makecell{\textbf{Complexity Level } \\ \textbf{(<word count)}}  & \textbf{Node F1 (\%)} & \textbf{Edge F1 (\%)} & \makecell{\textbf{Exact} \\ \textbf{Match (\%)}} \\
\hline
Simple (<8 )      & 100 & 100 & 100 \\
Moderate (<15)          & 99  & 93  & 96  \\
Complex (<25)          & 96  & 88  & 76  \\
Very Complex (26+)        & 87  & 69  & 31  \\
\hline
\end{tabular}
\end{table}

\subsection{Ablation Studies on Structural Components}

We analyze the contribution of structural graph extraction and staged verification to explain the observed performance gains.

\subsubsection{Structural Graph Extraction Accuracy}

We evaluate the quality of extracted structural dependency graphs by comparing them against ground-truth acquisition graphs recovered from passing programs via static AST analysis. Node and edge correspondences are resolved using an OpenDB API return-type table.

As shown in Fig.~\ref{causal_graph_acc}, the extracted graphs achieve strong node-level accuracy (F1 = 0.869), reliably recovering required object types. Recall (0.911) exceeds precision (0.861), indicating a tendency to add extra nodes rather than miss necessary ones. Edge-level accuracy is lower but still substantial (F1 = 0.680), and exact graph match is achieved in 25\% of cases \cite{sokolova2009systematic} .

Table~\ref{tab:complexity_accuracy} summarizes accuracy across prompt complexity. While performance remains near-perfect for simple tasks, exact match \cite{stengel2023calibrated} drops to 31\% for very complex prompts, confirming that precise structural recovery becomes increasingly difficult with complexity. Despite this, execution is not significantly impacted, as extra nodes improve retrieval coverage and missing dependencies are corrected during verification. Thus, approximate structural extraction provides a sufficient prior for correct execution .

\subsubsection{Verifier Contribution and Staged Enforcement}

Staged verification improves execution reliability mainly through structural constraint enforcement. Causal verification raises precision from 0.508 to 0.807 and F1 from 0.674 to 0.880 over static checking. Semantic checking then refines borderline cases, and the full staged verifier achieves the best overall precision-recall balance.

\begin{table}[t]
\centering
\caption{Verifier quality on single-task prompts. \vspace{-10pt}}
\label{tab:verifier_metrics}
\small
\begin{tabular}{lccc}
\toprule
Verifier Stage & Precision & Recall & F1 \\
\midrule
Static Verifier       & 0.508 & 1.000 & 0.674 \\
Causal Verifier       & 0.807 & 0.968 & 0.880 \\
LLM Semantic Verifier & 0.849 & 1.000 & \textbf{0.918} \\
Full Staged Verifier  & 0.849 & 0.909 & 0.878 \\
\bottomrule
\end{tabular}
\end{table}

\begin{table}[t]
\centering
\caption{Effect of uncertainty-aware filtering on pass rate and verifier precision.}
\label{tab:uncertainty_passrate}
\scriptsize
\setlength{\tabcolsep}{4pt}
\renewcommand{\arraystretch}{1.1}
\resizebox{\columnwidth}{!}{%
\begin{tabular}{l c c c}
\toprule
Metric & Baseline & + Uncertainty & $\Delta$ \\
\midrule
Step 1 pass rate             & 80.0\% & 92.0\% & +12.0\% \\
Step 2 pass rate             & 78.0\% & 90.0\% & +12.0\% \\
Overall pass rate            & 72.0\% & 84.0\% & +12.0\% \\
Verifier false positive rate & 20.0\% & 6.7\%  & $-13.3\%$ \\
Verifier precision           & 80.0\% & 93.3\% & +13.3\% \\
\bottomrule
\end{tabular}%
}
\end{table}

\begin{table}[t]
\centering
\caption{Cross-design and cross-PDK (Process Design Kit) generalization of the proposed framework. Inst: standard-cell instances; Nets: signal nets; IO: input/output pins; Lat (s): average task latency; LLM: average large language model calls per task; Cost: estimated cost per task using GPT-4.1-mini pricing (\$0.40/M input tokens, \$1.60/M output tokens).}
\label{tab:generalization}
\small
\setlength{\tabcolsep}{2pt}
\begin{tabular}{l l c r r r c r r r}
\toprule
Design & PDK & Node & Inst & Nets & IO & Pass & Lat (s) & LLM & Cost (\$) \\
\midrule
GCD    & Nangate45 & 45nm  & 624     & 581    & 54  & 81.5\% & 28.0 & 3.0 & 0.0028 \\
GCD    & sky130hd  & 130nm & 1,143   & 306    & 54  & 88.9\% & 20.0 & 2.2 & 0.0021 \\
GCD    & sky130hs  & 130nm & 366     & 402    & 54  & 94.4\% & 24.1 & 2.4 & 0.0019 \\
AES    & Nangate45 & 45nm  & 156,172 & 17,388 & 388 & 88.9\% & 51.5 & 2.1 & 0.0020 \\
RISC-V & ASAP7     & 7nm   & 12,161  & 9,585  & 135 & 90.7\% & 22.5 & 1.26 & 0.0017 \\
AES    & ASAP7     & 7nm   & 22,107  & 21,542 & 388 & 94.4\% & 21.6 & 1.04 & 0.0013 \\
\bottomrule
\end{tabular}
\end{table}

\subsection{Uncertainty and Reliability Analysis}

While staged verification substantially improves execution correctness (Section~4.2), residual failures arise when verification is incomplete or repair does not converge. We therefore introduce uncertainty measures to estimate reliability prior to execution and enable selective filtering of low-confidence programs. Table~\ref{tab:uncertainty_passrate} summarizes the impact of uncertainty-aware filtering. Without filtering, the verifier exhibits a false positive rate of 20.0\% and precision of 80.0\%, indicating that a non-trivial fraction of programs marked as valid still fail at execution. Incorporating uncertainty reduces the false positive rate to \textbf{6.7\%} and improves precision to \textbf{93.3\%}, showing that uncertainty signals effectively identify unreliable candidates.

As explained in the section 3.5 we consider three complementary sources of uncertainty \cite{wang2022self}.  These signals provide complementary views of failure: code-level uncertainty targets local inconsistencies, trajectory-level uncertainty reflects weak convergence, and coverage uncertainty captures blind spots in verification . Combining them yields a more informative reliability estimate than the verifier’s binary decision \cite{geifman2017selective}.

\begin{figure}[t]
    \centering
    \includegraphics[width=1\columnwidth]{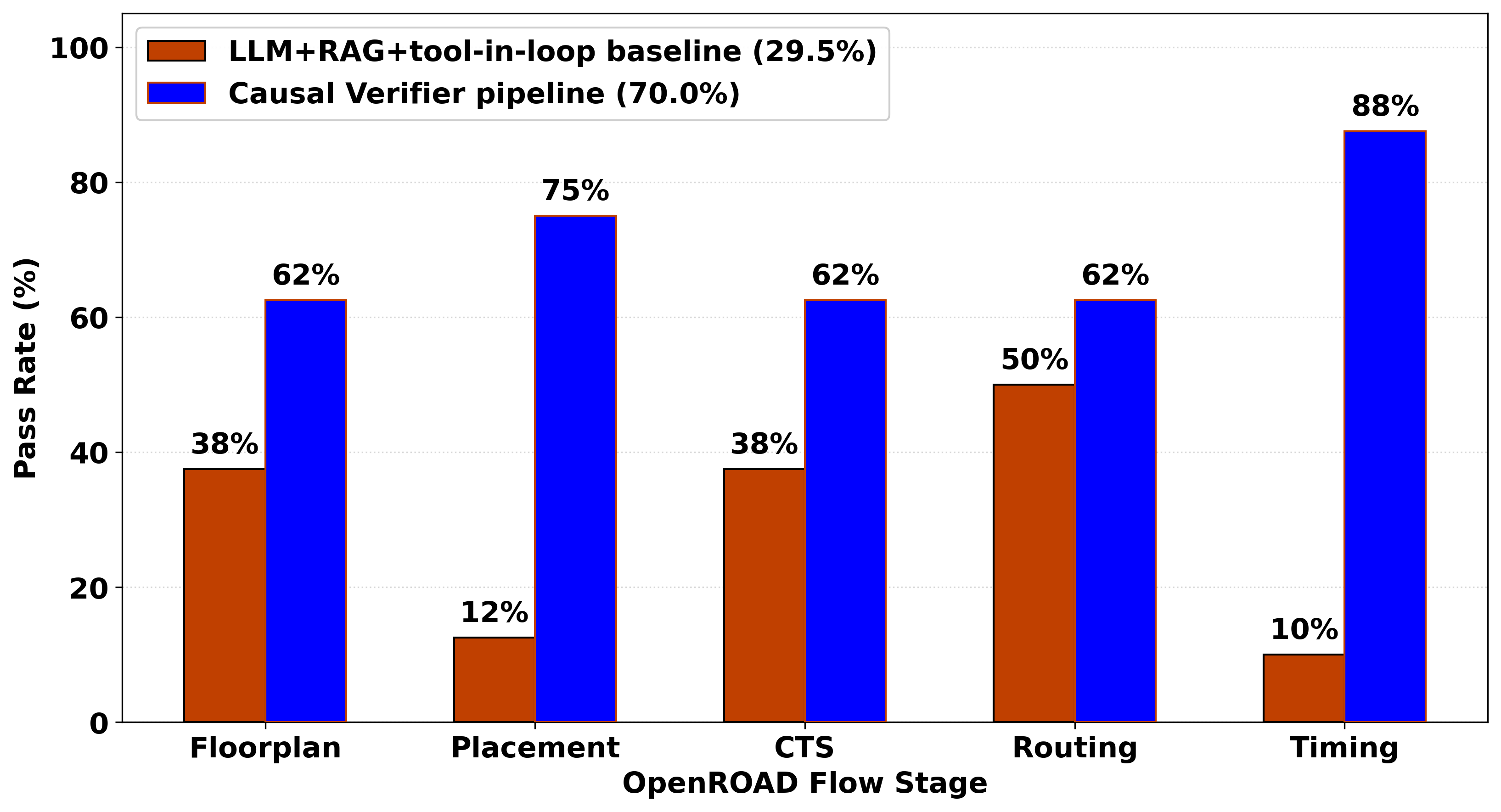}
    \caption{Stage-wise performance across OpenROAD flow stages, showing consistent pass rates across diverse operations and API patterns.}
    \label{fig:stage_generalization}
\end{figure}

\subsection{Generalization Across Designs and PDKs}

We evaluate the proposed framework across different designs, process design kits (PDKs), and OpenROAD flow stages. Unlike prompt-level approaches that may overfit to specific API patterns, structural verification operates on dependency constraints, enabling transfer across diverse configurations.

Table~\ref{tab:generalization} summarizes performance across designs and PDKs. The framework maintains high pass rates, indicating that structural consistency generalizes beyond prompt-construction distributions. Performance remains stable across variations in object hierarchies, API usage patterns, and flow stages. Fig.~\ref{fig:stage_generalization} shows stage-wise results across OpenROAD tasks, including floorplanning, placement, CTS, routing, and timing. The method achieves consistent accuracy across stages, demonstrating robustness to diverse operations and API compositions. In contrast, baseline approaches show higher variability. Retrieval-based methods are sensitive to shifts in API usage, while tool-in-the-loop approaches incur higher cost from repeated execution, especially when errors propagate across stage-specific dependencies. These results suggest that structural dependency modeling captures invariant properties of EDA workflows, improving robustness while maintaining low tool usage.

\section{Conclusion}

We presented a framework that eliminates tool-in-the-loop debugging in LLM-based EDA code generation by enforcing structural correctness before execution. By modeling tasks as structural dependency graphs used as execution contracts, the approach shifts correctness from post-hoc feedback to pre-execution verification, enabling graph-conditioned synthesis, staged verification, and diagnosis-driven repair with a single tool call per task. Experiments in OpenROAD show improved reliability and efficiency. The method reaches 82.5\% single-step pass rate while reducing tool calls by more than 2$\times$, and scales to multi-step tasks, achieving up to 84.0\% with trajectory-level reflection. Uncertainty-aware filtering further improves reliability, and results across designs, PDKs, and flow stages show strong generalization. Overall, the results suggest that failures in EDA code generation are fundamentally structural, and that enforcing dependency consistency before execution offers a scalable alternative to execution-driven correction.

\bibliographystyle{ACM-Reference-Format}
\bibliography{main}
\end{document}